\documentstyle[12pt]{article}

\def\hybrid{\topmargin -20pt  \oddsidemargin 0pt
      \headheight 0pt   \headsep 0pt
      \textwidth 6.25in 
      \textheight 9.5in 
      \marginparwidth .875in
      \parskip 5pt plus 1pt   \jot = 1.5ex}

\hybrid

\begin{document}
\def\x{\times}
\def\beq{\begin{equation}}
\def\eeq{\end{equation}}
\def\beqa{\begin{eqnarray}}
\def\eeqa{\end{eqnarray}}

\sloppy
\newcommand{\be}{\begin{equation}}
\newcommand{\eq}{\end{equation}}
\newcommand{\ov}{\overline}
\newcommand{\un}{\underline}
\newcommand{\p}{\partial}
\newcommand{\la}{\langle}
\newcommand{\ra}{\rangle}
\newcommand{\bl}{\boldmath}
\newcommand{\ds}{\displaystyle}
\newcommand{\nl}{\newline}
\newcommand{\th}{\theta} 
\newcommand{\resetcounter}{\setcounter{equation}{0}}     
\newcommand{\non} {\nonumber}             
\newcommand{\ga}  {\gamma}                  
\newcommand{\GA}  {\Gamma}                  
\newcommand{\ti}  {\tilde}                  
\newcommand{\T}  {\theta}
\newcommand{\Tb} {\bar\theta}
\newcommand{\A}  {\alpha}
\newcommand{\B}  {\beta}
\newcommand{\E}  {\varepsilon}
\newcommand{\ph}  {\varphi}
\newcommand{\de}  {\delta}
\newcommand{\ze}  {\zeta}
\newcommand{\si}  {\sigma}
\newcommand{\sd} {\sigma^{m} \partial_{m}}
\newcommand{\tsdt} {\theta \sigma^{m} {\bar\theta} \partial_{m}}
\newcommand{\tst} {\theta \sigma^{m} {\bar\theta}}
\newcommand{\Sb} {\bar\sigma}
\newcommand{\C} {\chi}
\newcommand{\Cb} {\bar\chi}
\newcommand{\Ps}  {\psi}
\newcommand{\Pb} {\bar\psi}
\newcommand{\UL} {\underline}
\newcommand{\G}  {{\cal G}}   
\newcommand{\R}  {{\dot{R}}}   %
\chardef\ss='031


\renewcommand{\thesection}{\arabic{section}}
\renewcommand{\theequation}{\thesection.\arabic{equation}}

\parindent0em

\begin{titlepage}
\begin{center}
\hfill HUB-EP-97/43\\
\hfill {\tt hep-th/9707243}\\

\vskip .7in  

{\bf  A NOTE ON A CLASS OF COSMOLOGICAL \\ STRING BACKGROUNDS}

\vskip .3in

Bj\"orn Andreas\footnote{email: andreas@qft3.physik.hu-berlin.de}
\\
\vskip 1.2cm
{\em Humboldt-Universit\"at zu Berlin,
Institut f\"ur Physik, 
D-10115 Berlin, Germany}

\vskip .1in

\end{center}

\vskip .2in

\begin{quotation}\noindent

We study a class of four dimensional, 
anisotropic string backgrounds  
and analyse their expansion and singularity structure.
In particular we will see how O(3,3) duality acts 
on this class.

\end{quotation}
\end{titlepage}
\vfill
\eject

\newpage

In the course of study of string cosmological backgrounds one class
of solutions appeared over the time which was origanally introduced 
by Mueller [\ref{mue}] and was studied in the string frame 
[\ref{lue},\ref{tsy}] and in related works [\ref{rela}].

In this note we will study the space of solutions in the Einstein
frame which is related to the string frame by a conformal transformation.
We will show that if one introduces a new time coordinate within the
Einstein frame one can apply techniques known from General Relativity
to get a geometrical interpretation of the expansion and singularity 
structure of the solutions. Furthermore we will see how O(3,3) duality 
acts on this space and it will be shown that depending on the anisotropy 
of the solution different types of singularities appear.\\

Now let us start and consider a string propagating in the presence 
of a D-dimensional background spacetime metric $G(X)$,  
a dilaton $\Phi(X)$ and the antisymmetric tensor field $B(X)$ with 
field strenght $H(X)$. The string tree level effective action for these 
fields is given in the Einstein frame by 
\begin{eqnarray} 
S=\int{d^Dx}\sqrt{-G}[R-2{(\nabla\phi)}^2-
e^{-4\phi}\frac{1}{12}H^2]+S_{f}.\non
\end{eqnarray}
where $S_{f}$ denotes that there can be moduli fields coming from 
compactification and higher loop effects, etc. By varying $G,B,\phi$ we 
obtain the equations of motion 
\begin{eqnarray} 
R_{\mu\nu}-\frac{1}{2}G_{\mu\nu}R &=& T_{\mu\nu}\non\\
\nabla_{\mu}(e^{-4\phi}H^{\mu}_{\nu\lambda}) &=& 0\non\\
\nabla^2\phi + e^{-4\phi}\frac{1}{12} H^2 &=& 0 .\non
\end{eqnarray}
with the energy-momentum tensor for the axion-dilaton system given by  
\begin{eqnarray} 
T_{\mu\nu} &=& \frac{1}{4}(H_{\mu\nu}^2-\frac{1}{6}G_{\mu\nu}H^2)e^{-4\phi}
        +2(\nabla_{\mu}\phi\nabla_{\nu}\phi -\frac{1}{2}G_{\mu\nu}(\nabla
        \phi)^2 )\non
\end{eqnarray}
These equations can also be obtained [\ref{gre},\ref{cal}] by the 
conditions of
vanishing $\beta$-functions in the corresponding two-dimensional $\sigma$
model.\\ 
Now let us consider as an ansatz the anisotropic spacetime metric
which has a time dependent dilaton and a spacetime metric given 
in the Einstein frame by (setting $B(X)=0$)
\begin{eqnarray}
ds_{E}^2 = -e^{-2\phi}dt^2 + \sum_{i=1}^3({R}_{i}(t))^2(dX^i)^2,\non
\end{eqnarray}
where $R_{i}(t)^{2}$ is given by $R_{i}(t)^{2}=e^{-2\phi}{R}_{i}^S(t)^{2}$
and the corresponding field equations (cf. for the stringframe 
[\ref{mue},\ref{lue}]) are
\begin{eqnarray}
\sum_{i<j}\frac{\R_{i}\R_{j}}{ R_{i} R_{j}}&=&(\dot{\phi}(t))^2 \non \\
-e^{2\phi} R_{i}^2\left [\frac{\R_{k}\R_{l}}{ R_{k} R_{l}}
+  \frac{\R_{k}}{ R_{k}}\dot{\phi}
+  \frac{\R_{l}}{ R_{l}}\dot{\phi}
+  \frac{\ddot { R}_{k}}{ R_{k}}
+  \frac{\ddot { R}_{l}}{ R_{l}}\right] &=& e^{2\phi} R_{i}(t)^2
(\dot{\phi}(t))^2.\non 
\end{eqnarray}
and $ i,k,l \in$ \{1,2,3\} with $k,l \neq i$ ($k>l$) further
$R_{i}=R_{i}(t)$ and $\phi=\phi(t)$. The dot denotes derivative with
respect to $t$.\\
A family of solutions introduced originally in the
string frame [\ref{mue}] are given in the Einstein frame by
\begin{eqnarray}
R_{i}(t) = \A_{i}(t-t_{0})^{p_{i}+\frac{p}{2}}\label{rad}\  \  \  \
\phi(t) = -\frac{1}{2}log\ \beta^2(t-t_{0})^p \label{phi}\non,  
\end{eqnarray}
where  $\A_{i}, \beta, t_{0}$ are arbitrary  real numbers;
$p_{i}$ and $p$ are real but restricted through
\begin{eqnarray}
\sum_{i=1}^3 p_{i}^2 = 1,\      \   \    \ \sum_{i=1}^3 p_{i} = 1-p, 
\      \   \    \sum_{i<j}p_{i}p_{j} &=& \frac{1}{2}(p^2-2p).\non
\end{eqnarray}
Now these equations lead to the geometrical picture that the 
$p_{i}$ can be thought of being points on the unit 2-sphere (note that 
$\sum_{i}p_i=1-p$ determines a codimension one subvariety in this 
sphere) and every choice of a tupel ($p_{i}$) determines a solution 
of the differential equation system and restricts the range of $p$. 
The solutions depend on 7 parameters :$\alpha_{i},\beta,t_0$ 
(fixing the initial values of $R_{i}$ und $\phi$) and 2 free coordinates 
on the 2-sphere. The behavior of the dilaton field depends on whether 
$p=0$, $p<0$ or $p>0$. If we consider the intersection set (which
is either one dimensional or empty) of the hyperplane
$\sum_{i=1}^{3}p_{i}+p-1=0$ with the sphere $\sum_{i=1}^{3}p_{i}^2-1=0$
we can further restrict the range of $p$. If the hyperplane is tangent 
to the sphere then there are two critical values of $p$ which occur
for $1-p=\pm \sqrt{3}$ and thus the range of $p$ is      
($1-\sqrt{3}\le p \le 1+\sqrt{3}$).\\
Now to get rid of the scale factor in front of our line element 
let us introduce a new time coordinate $\tau$ that leads to a
line element which can be studied by known methods from
General Relativity.  
\begin{eqnarray}
ds_{E}^{2}= -d{\tau}^{2}+\sum_{i}^{3}R_{i}(\tau)^{2}d(X^{i})^2 . 
\non
\end{eqnarray}
The corresponding coordinate transformation can be obtained from 
$\dot{t}(\tau)=t(\tau)^{-\frac{p}{2}}$ where the dot denotes now and
in the following the derivative with respect to $\tau$. A solution is
given by 
\begin{eqnarray}
{t}(\tau)=(\tau+\frac{p}{2}\tau+c)^{\frac{1}{1+\frac{p}{2}}},\ \ c=const.\ .
\non
\end{eqnarray}
The solutions  are given in the new time coordinate by 
\begin{eqnarray}
R_{i}(\tau) = t(\tau)^{p_{i}+\frac{p}{2}},\      \   \    \
\phi(\tau) = -\frac{1}{2}ln\ t(\tau)^p \non 
\end{eqnarray}
where we have set for simplicity $\A_{i},\beta=1$ and $t_{0}=0$.\\ 
If we insert our new line element into the equations of motion 
of our background fields we find
\begin{eqnarray} 
\sum_{i<j}\frac{\R_{i}\R_{j}}{ R_{i}R_{j}} &=& p^{2}
\frac{\dot{t}(\tau)^{2}}{4t(\tau)^{2}}\non\\
-R_{i}^2\left [\frac{\R_{k}\R_{l}}{ R_{k} R_{l}}
+  \frac{\ddot{R}_{k}}{R_{k}}
+  \frac{\ddot{R}_{l}}{R_{l}}\right] &=& p^{2}\frac{t(\tau)^{2p_{i}+p}
\dot{t}(\tau)^{2}}{4t(\tau)^{2}}\non
\end{eqnarray}
with $i,k,l\in \{1,2,3\}$ and $k,l\neq i$ and $k>l$. The energy density
$\mu$ and the pressure $\hat{p}$ of the dilaton matter system are 
given by 
\begin{eqnarray}
T_{00} =  \mu(\tau), \  \  \  \ T_{ij} = \hat{p}(\tau)G_{ij}
=p^{2}\frac{\dot{t}(\tau)^{2}}{4t(\tau)^{2}}G_{ij}.\non
\end{eqnarray}
Now to decide which behavior our solutions have (i.e. if they describe
a contracting or expanding universe and if there is an initial singularity) 
we have to consider the energy-momentum conservation which leads to the 
following equation
\begin{eqnarray}
D^{\mu}T_{\mu\nu} &=& G^{\mu\lambda}\partial_{\lambda}T_{\mu\nu}
                      -G^{\mu\nu}\GA_{\mu\lambda}^{\rho}T_{\rho\nu}
                      -G^{\mu\nu}\GA_{\nu\lambda}^{\rho}T_{\mu\rho}\non\\
                  &=& -\dot{\mu}-({\mu}+\hat{p})
                        \sum_{i}\frac{\dot{R}_{i}(\tau)}{R_{i}(\tau)}\non\\
                  &=& 0 \non
\end{eqnarray}
Since $\frac{\R_{i}}{R_{i}}$ measures the expansion respectively 
contraction of the universe with respect to the i-th coordinate, we learn
\begin{eqnarray}
\frac{\R_{i}}{R_{i}}>0\ \ \Longleftrightarrow {\rm Expansion};\  \  \  \
\frac{\R_{i}}{R_{i}}<0\ \ \Longleftrightarrow {\rm Kontraktion}\non 
\end{eqnarray}
Finally we have to study the Raychandhuri equation in order to decide
if we have an initial singularity [\ref{haw}]. This equation can be 
derived if one considers the energy-momentum tensor of our dilaton matter
system which can be thought of as a  perfect fluid and is given by
\begin{eqnarray}
T_{\mu\nu}=\hat{p}G_{\mu\nu}+(\mu+\hat{p})V_{\mu}V_{\nu}\non.
\end{eqnarray}
where $V_{\mu}$ is the tangential vector to a congruence of timelike 
curves (with $g({\bf V},{\bf V})=-1$). Now every congruence of timelike
geodesics converges if $R_{\mu\nu}V^{\mu}V^{\nu}\ge0$ [\ref{haw}]. The 
equations of motion satisfies this condition if the energy-momentum tensor
satisfies  
\begin{eqnarray}
T_{\mu\nu}V^{\mu}V^{\nu}-\frac{1}{2}V^{\mu}V_{\mu}T\ge0,\ \ \ \
\mu+\hat{p}_{i}\ge0,\ \ \ \ \mu+\sum_{i}\hat{p}_{i}\ge0.\non
\end{eqnarray}
where $T=Tr\ \ T_{\mu\nu}$. This condition will be satisfied if the 
energy-momentum tensor is diagonal  $T_{\mu\nu}=diag(\mu,\hat{P}_{i})$
with $\hat{P}_{i}=\hat{p}_{i}G_{ij}$; the left hand side of this equation  
is the source term in the contracted Einstein equations  
\begin{eqnarray} 
R_{\mu\nu}=-8\pi S_{\mu\nu},\non
\end{eqnarray}
where $S_{\mu\nu}=T_{\mu\nu}+\frac{1}{2}G_{\mu\nu}T=\frac{1}{2}
(\mu-\hat{p})+(\hat{p}+\mu)V_{\mu}V_{\nu}$. Here we used  
the energy-momentum tensor. With our above ansatz $S_{\mu\nu}$ reduces to
\begin{eqnarray} 
S_{00} = \frac{1}{2}(\mu+3\hat{p}),\  \  \
S_{0i} = 0,\  \  \
S_{ij} = \frac{1}{2}(\mu-\hat{p})R_{i}(\tau)\Omega_{ij}(X)\non
\end{eqnarray}
where $\Omega_{ij}(X)$ is the metric of the threedimensional area of constant
curvature. The time-time component of $R_{\mu\nu}$ is then
\begin{eqnarray} 
R_{00}=4\pi(\mu+3\hat{p}).\non
\end{eqnarray}
and if we use $R_{00}=-\sum_{i}\frac{\ddot{R}_{i}(\tau)}{R_{i}(\tau)}$
we find the Raychandhuri equation \footnote{   
$\frac{\ddot{R}_{i}}{R_{i}}$ measures the acceleration
of the expansion resp. contraction and implies
\begin{eqnarray}
\sum_{i}
\frac{\ddot R_{i}}{R_{i}}<0\ \ \Longleftrightarrow \mu+3\hat{p}>0;\  \  \
\sum_{i}
\frac{\ddot R_{i}}{R_{i}}>0\ \ \Longleftrightarrow \mu+3\hat{p}<0.\non
\end{eqnarray}}
\begin{eqnarray} 
\sum_{i}\frac{\ddot{R}_{i}(\tau)}{R_{i}(\tau)}=-4\pi(\mu+3\hat{p}).\non
\end{eqnarray}
From the energy-momentum conservation we learn that the density of 
the universe vanishes for $R(\tau)\rightarrow \infty$ and grows up 
to infinity for $R(\tau)\rightarrow 0$. Since the density is a scalar
we can conclude that also the scalar curvature grows up and at
$R=0$ we have a singularity which is no coordinate singularity and leads
usally to the break down of the Einstein theory. Thus from the
Raychandhuri equation we learn that if $\mu+3\hat{p}>0$ we have an initial
singularity [\ref{haw}].\\ 
Now we are able to study the structure of our model where we have  
\begin{eqnarray}
R_{00}= \frac{p^{2}\dot{t}(\tau)^{2}}{2 t(\tau)^{2}}, \  \  \  \    
R_{ii} = 0, \  \  \  \
R=-\frac{p^{2}\dot t(\tau)^{2}}{2 t(\tau)^{2}}, \  \  \  \    
\mu=\hat{p}=\frac{p^{2}\dot{t}(\tau)^{2}}{4t(\tau)^{2}}.\non
\end{eqnarray}
{\underline{\bf Duality transformation}}\\ 

Since our model is independent of three coordinates we can perform 
a O(3,3) duality transformation (for a review on O(d,d) duality see
[\ref{giv}] and references therein). We do this in the string frame
which is related to the Einstein frame by the conformal transformation
$G_{\mu\nu}^S = e^{2\phi}G_{\mu\nu}^E$ and then transform 
back to the Einstein frame and find the dual line element
\begin{eqnarray}
ds_{D}^{2}=-d{\tau}^{2}+\sum_{i}(R_{i}^{D})^{2}(dX^{i})^2,\non
\end{eqnarray}
where $d{\tau}^{2}=e^{-2{\phi}^D}d{\bar{t}}^2$ and $(R_{i}^{D})^{2}=
e^{-2{\phi}^D}\ti{R}_{i}^{-2}$. The equations of motion are solved by
\begin{eqnarray}
R_{i}(\tau)^{D} = \bar{t}(\tau)^{-p_{i}-\frac{p}{2}+1},\  \  \  \
\phi(\tau)^{D} = -\frac{1}{2}log\ \bar{t}(\tau)^{2-p}.\non
\end{eqnarray}
where $\bar{t}(\tau)$ is given by rescaling of the dual metric.
The duality transformation $p_{i}\longrightarrow -p_{i}$,
$p\longrightarrow -p+2$ or equivalently  
$R_{i}\longrightarrow \frac{1}{R_{i}},\ \ \ p\longrightarrow -p+2,\ \ \ 
(i=1,2,3)$ relates the two string backgrounds to each other. 
Remember the range 
of $p$ was given by $p_{\rm min}=1-\sqrt{3}\le p \le 1+\sqrt{3}=p_{\rm max}$,
and therefore $p\rightarrow -p+2$ maps $p_{\rm min}\rightarrow 
p_{\rm max}$ resp. $p_{\rm max}\rightarrow p_{\rm min}$, i.e. the 
duality transformation operates on the space of solutions. 
For the dual model we find   
\begin{eqnarray}
R_{00}^{D}&=& \frac{(-p+2)^{2}\dot{\bar{t}}(\tau)^{2}}{2 \bar{t}
(\tau)^{2}}, \ \ \ 
R_{ii}^{D} = 0,\ \ \ {\bar{t}}(\tau)=(\tau(2-\frac{p}{2})+c)^{\frac{1}
{2-\frac{p}{2}}}\non \\
R^{D}&=&-\frac{(-p+2)^{2}\dot {\bar{t}}(\tau)^{2}}
{2 \bar{t}(\tau)^{2}},\ \ \   
\mu_{D}=\hat{p}_{D}=\frac{(-p+2)^{2}\dot{\bar{t}}(\tau)^{2}}{4\bar{t}
(\tau)^{2}}\non  
\end{eqnarray}
{\underline{\bf Singularities}}\\ 

Let us study the singularity structure. Setting $c=0$ we obtain (using
$\frac{\dot{t}(\tau)^{2}}{t(\tau)^{2}}= (\tau(1+\frac{p}{2}))^\frac{-2p-4}
{2+p}\label{tau}$ and $\frac{\dot{\bar{t}}(\tau)^{2}}{\bar{t}(\tau)^{2}}=
(\tau(2-\frac{p}{2}))^{\frac{2p-8}{-p+4}}$) for the
scalar curvatures $R$, $R^{D}$ and the energy densities $\mu$, $\mu^{D}$
\begin{eqnarray}
R &=& -\frac{p^2}{2}(\tau(1+\frac{p}{2}))^\frac{-2p-4}{2+p},\  \  \
\mu = \frac{p^2}{4}(\tau(1+\frac{p}{2}))^\frac{-2p-4}{2+p}, \non\\
R^{D} &=& -\frac{(2-p)^2}{2}(\tau(2-\frac{p}{2}))^{\frac{2p-8}{-p+4}},\  \  \
\mu^{D} = \frac{(2-p)^{2}}{4}(\tau(2-\frac{p}{2}))^{\frac{2p-8}{-p+4}}.\non  
\end{eqnarray}
\underline{$p\neq 0$}:

For solutions with $p\neq 0$ both space-times are singular at $\tau=0$
which can be checked by inserting $p_{max}$ and $p_{min}$ into the
equations above. Furthermore for all $p\neq 0$ we have
\begin{eqnarray}
\mu+3\hat{p}>0, \  \  \  \mu^{D}+3\hat{p}^{D}>0\non     
\end{eqnarray}
and the energy density has the $\frac{1}{\tau^{a}}$ behavior typical 
for cosmological solutions and will grow up to infinity if 
$\tau\rightarrow 0$.

\underline{$p= 0$}: 

If $p=0$ then the dilaton is constant and the energy density
vanishes. The scalar curvature and energy density are given by 
\begin{eqnarray}
R=0,\ \ \ \mu=0,\ \ \  
R^{D}=-\frac{1}{2\tau^{2}},\ \ \ \mu^{D}=\frac{1}{\tau^2}.\non
\end{eqnarray}
Further $\mu+\hat{p}$ vanishes for Ricci-flat spacetimes. To get 
some information about singularities one has to check if the 
Riemannian tensor gets singular for $\tau\rightarrow 0$. In our case
we find $ R_{\mu\nu\lambda\si}R^{\mu\nu\lambda\si}\sim \frac{1}{\tau^2}$ 
which will be singular at $\tau=0$. Thus the duality transformation 
maps a Ricci-flat spacetime onto spacetime with singular scalar 
curvature.

{\underline{\bf Expansion behavior}}\\ 

To understand the expansion behavior we have to 
consider $\frac{\dot{R}_{i}}{R_{i}}=\frac{p_{i}+\frac{p}{2}}{1+\frac{p}{2}}
\left(\frac{1}{\tau}\right)$ and $\frac{\dot{R}_{i}^{D}}{R_{i}^{D}}=
\frac{-p_{i}-\frac{p}{2}+1}{2-\frac{p}{2}}\left(\frac{1}{\tau}\right)$.

\underline{$p= 0$}:\\ 

The Expansion behavior for $p=0$ is simply given by
\begin{eqnarray}
\frac{\dot{R}_{i}}{R_{i}}=\frac{p_{i}}{\tau}\ \ \ \ , \ \ \ \ 
\frac{\dot{R}_{i}^{D}}{R_{i}^{D}}=
\frac{-p_{i}+1}{2\tau}.\non
\end{eqnarray}
Now remember $p_{i}$ are coordinates of points on the unit sphere 
with values in [-1,1]. Since our solutions are 4-dimensional we exclude
cases with $p_{1}=1,p_{i}=0$ $\{i=2,3\}$ and $p_{i}=0$ $\{i=1,2,3\}$.
Taking into acount that the dual expansion coefficient being positive
for all $p_{i}$ we learn that our dual solution describes an expanding 
spacetime. The original solution describes also an expanding spacetime 
if $p_{i}$ is positive with respect to the i-th coordinate and describes
an contracting spacetime if $p_{i}$ is negative with respect to the 
i-th coordinate 
\begin{eqnarray}
p=0:\ \ \ p_{i} \left\{ \begin{array}{r@{\quad: \quad}l}
                   >0&  {\rm EXPANSION\rightarrow EXPANSION}\non\\
                   <0&  {\rm CONTRACTION\rightarrow EXPANSION} 
                                      \end{array} \right.
\end{eqnarray}

If we solve the equation system $\sum_{i}p_{i}=1, \ \ \sum_{i}p_{i}^{2}=1$
we will find coordinates for $p_{i}$. For 
example if we fix $p_{3}$ then we will obtain  $p_{1}$ and $p_{2}$:
\begin{eqnarray}
p_{1} = \frac{1}{2}\left(1-p_{3}\mp\sqrt{1+2p_{3}-3p_{3}^{2}}\right),\ \ \
p_{2} = \frac{1}{2}\left(1-p_{3}\pm\sqrt{1+2p_{3}-3p_{3}^{2}}\right).\non
\end{eqnarray}
\underline{Example}: 
$p = 0, \ \ \ p_{3}=\frac{1}{2}>0, \ \ \ p_{1}= \frac{1}{4}-
\frac{\sqrt{5}}{4}<0,\ \ \ p_{2}=\frac{1}{4}+\frac{\sqrt{5}}{4}>0$

In the example duality maps a Ricci-flat spacetime which expands in
$i=3,2$ and contracts in  $i=1$ to a spacetime which expands  
in $i=1,2,3$. Both spacetimes are singular for $\tau\rightarrow 0$ 
but with different singularity structure. The former has an infinite 
THREAD singularity if $\tau\rightarrow 0$ and the latter a POINT 
singularity. 

\underline{$p\neq 0$}:\\ 

Now let us come to the case $p\neq 0$ where the range for $p$ and
$p_{i}$ is $1-\sqrt{3} \le p \le 1+\sqrt{3}$ and $ -1<p_{i}<1$ respectively. 
Since the denomintator of $\frac{\dot{R}_{i}}{R_{i}}$ and
$\frac{\dot{R}_{i}^{D}}{R_{i}^{D}}$ is always positive we are left
to check the nominator. We find that we can distinguish the following cases: 

A:
\underline{STATIC$\longrightarrow$ EXPANSION:}
\begin{eqnarray}
\left(p_{i}+\frac{p}{2}\right)=0 \longrightarrow 
-\left(p_{i}+\frac{p}{2}\right)+1>0\non 
\end{eqnarray}
B:
\underline
{CONTRACTION$\longrightarrow$ EXPANSION:}
\begin{eqnarray}
\left(p_{i}+\frac{p}{2}\right)<0 \longrightarrow 
-\left(p_{i}+\frac{p}{2}\right)+1>0\non
\end{eqnarray}
C:
\underline
{EXPANSION$\longrightarrow$ EXPANSION:}
\begin{eqnarray}
0<\left(p_{i}+\frac{p}{2}\right)<1 \longrightarrow 
-\left(p_{i}+\frac{p}{2}\right)+1>0\non 
\end{eqnarray} 
D:
\underline
{EXPANSION$\longrightarrow$ CONTRACTION:}
\begin{eqnarray}
\left(p_{i}+\frac{p}{2}\right)>1 \longrightarrow 
-\left(p_{i}+\frac{p}{2}\right)+1<0.\non
\end{eqnarray}

{\underline{\bf Conclusion}}\\ 

In this note we showed that one can get a geometrical interpretation 
of the space of solutions which is parametrised by $p$ and $p_{i}$ using 
methods known from classical General Relativity. 
The duality transformation $p_{i}\rightarrow -p_{i}, p\rightarrow
-p+2$ acting on this space maps for $p=0$ and $p_{i}>0$
expanding to expanding backgrounds and for $p=0$ and $p_{i}<0$
contracting to expanding backgrounds and if 
$p\neq0$ the cases A-D appear. Furthermore it has been shown that
depending on the possible combinations of A-D there are backgrounds
with THREAD singularities as well as POINT or PANCAKE singularities 
as a consequence of having an anisotropic spacetime. 
\vskip0.5cm

{\bf Acknowlegement:}
I would like to thank C.Curio and D.L\"ust for discussions.
The work has been supported by NAFOEG.

\section*{References}
\begin{enumerate}
\item
\label{mue}
M. Mueller Nucl.Phys.{\bf B337}(1990) 37;

\item
\label{lue}
D. L\"ust, preprint CERN-TH.6850/93; 

\item
\label{tsy}
A.A. Tseytlin and C. Vafa Nucl.Phys.{\bf B372}(1992) 443;

\item
\label{rela}
E. Kiritsis, C. Kounnas and D. L\"ust
Int.J.Mod.Phys.{\bf A9}(1994) 1361; I. Bakas Nucl.Phys.{\bf B428}(1994) 374;
I. Antoniades, S. Ferrara and C. Kounnas  Nucl.Phys.{\bf B421}(1994) 343;
M. Gasperini and G. Veneziano Phys.Rev.{\bf D50}(1994) 2519; 
A.A. Tseytlin Phys.Lett.{\bf B334}(1994) 315; M. Gasperini and R. Ricci 
Class.Quant.Grav.{\bf 12}(1995) 677; N.A. Batakis and  A.A. Kehagias 
Nucl.Phys.{\bf B449}(1995) 248; A. Kehagias and A. Lukas    
Nucl.Phys.{\bf B477}(1996) 549; A. Lukas, B. A. Ovrut and D. Waldram  
Phys.Lett.{\bf B393}(1997) 65; A. Lukas, B. A. Ovrut and D. Waldram  
Nucl.Phys.{\bf B495}(1997) 365; E. Kiritsis and C. Kounnas gr-qc/9701005
publ.in Nucl.Phys.Proc.Suppl.{\bf 41}(1995) 365; J.D. Barrow and
K.E. Kunze Phys.Rev.{\bf D56}(1997) 741; M.P. Dabrowski, A.L. Larsen
hep-th/9706020;

\item
\label{gre}
M.B.Green, J.Schwarz and E.Witten: Superstring theory
Cambridge University Press, 1987;

\item
\label{cal}
C.G.Callan, D.Friedan, E.J.Martinec und M.J.Perry, Nucl.Phys.{\bf B262}
,593;

\item
\label{haw}
S.W.Hawking und G.F.R.Ellis: The large scale structure of space-time,
Cambridge Monographs on Mathematical Physics;

\item
\label{giv}
A. Giveon, M. Porrati and E. Rabinovici Phys.Rept.{\bf 244}(1994) 77;
\end{enumerate}

\end{document}